# Lowest-energy structures of 13-atom binary clusters: Do icosahedral clusters exist in binary liquid alloys?

M. Iwamatsu[a,*], S. K. Lai[b]

[a]Department of Physics, General Education Center, Musashi Institute of Technology,

Setagaya-ku, Tokyo 158-8557, Japan

[b]Complex Liquids Laboratory, Department of Physics, National Central University,

Chung-li 320, Taiwan, Republic of China

**Abstract**

Although the existence of 13-atom icosahedral clusters in one-component close-packed undercooled liquids was predicted more than half a century ago by Frank, the existence of such icosahedral clusters is less clear in liquid alloys.  We study the lowest-energy structures of 13-atom $A_xB_{13-x}$ Lennard-Jones binary clusters using the modified space-fixed genetic algorithm and the artificial Lennard-Jones potential designed by Kob and Andersen.  Curiously, the lowest-energy structures are non-icosahedral for almost all compositions.  The role played by the icosahedral cluster in a binary glass is questionable.

## 1. Introduction

The existence of 13-atom icosahedral clusters in close-packed undercooled metallic liquids was predicted more than half a century ago by Frank [1] from his total energy calculation of isolated clusters. He compared the total potential energy of a 13-atom regular icosahedral cluster with those of 13-atom FCC (cubo-octahedral) and 13-atom HCP (hexagonal close-packing) clusters, and found that the former has a lower energy than the latter. His choice of the 13-atom cluster is also of interest for the well-known fact that the atoms in a close-packed liquid are surrounded on average by 12 atoms.

---

[*] Corresponding author. Tel: +81-3-3703-3111; fax: +81-3-5707-2222;
e-mail:iwamatsu@ph.ns.musashi-tech.ac.jp

Because of the stability of this regular icosahedron over the standard crystallographic FCC or HCP structures, the icosahedral cluster is considered as a basic building block of amorphous (amorphons) metals [2,3]. Here, the icosahedral cluster means not only the regular but also the distorted icosahedron as we are interested in disordered liquid and amorphous structures. The five-fold symmetry of the icosahedral arrangement of atoms known as the bond-orientation order [4], which is incompatible to the long-ranged translational symmetry, has long been considered as the essential ingredient of metallic glasses [5,6]. Subsequent computer simulations [7], and experiments by neutron scattering [8], X-ray scattering [9] and X-ray absorption [10] clearly indicated the existence of the distorted icosahedral clusters in one-component liquids and glasses.

In liquid alloys, however, the existence of such 13-atom icosahedral clusters is less clear though the special role played by the polytetrahedral order in complex solid alloys [11] seems to suggest the stability of icosahedral clusters in undercooled liquid alloys as well. However, the connection between the glass formation and the stability of the icosahedral cluster or the five-fold symmetric bond is not well understood for alloys even though various experimental evidences from neutron scattering [12] or X-ray absorption fine structure (EXAFS) [13] as well as the theoretical evidence from simulations [14,15] have been accumulated.

In our report, we follow the strategy taken by Frank [1], and determine the lower-energy structures of isolated 13-atom $A_xB_{13-x}$ Lennard-Jones binary clusters. A similar study has been already conducted by Cozzini and Ronchetti [16] using molecular dynamics. We use, instead, the modified space-fixed genetic algorithm developed for multi-component systems [17] in this work. Although the stability of a cluster in liquid and free space would be different, this problem was partly answered using a mean-field like theory by Mossa and Tarjus [18]. They took into account the potential made by the surrounding atoms in liquid, and found that the liquid-like environment only slightly affects the stability of the cluster.

## 2. Modified space-fixed genetic algorithm

Our genetic algorithm (GA) is basically the same as the one proposed previously [17], which was used to study the $Ar_xXe_{13-x}$ binary clusters. The genetic algorithm is a method to search for the lowest-energy structure of clusters. We prepare a population that consists of $N$ clusters, which are called individuals each consists of $x$ A atoms and 13-$x$ B atoms. In a GA, each of the clusters produces child structures called the next generation. The four genetic operations [17] called: Inversion, Arithmetic mean, Geometric mean, $m$-points crossover, are used to produce new structures. Subsequently, a simplex minimization [19] is performed for each new individual in order to place each cluster at its local minimum. By using this parallel search and the information exchange between searches, the GA can efficiently locate the global minimum of complex multidimensional functions. The detail of the algorithm can be found in reference [17]. In this work, however, a special care concerning the final solution was taken by performing multiple trials of simplex minimization in order to guarantee the convergence since we are not only interested in the lowest-energy structure but also the higher-energy structures.

## 3. Numerical results

We studied the lower energy structures of $A_xB_{13-x}$ Lennard-Jones binary clusters using the modified space-fixed genetic algorithm. We chose four genetic operators randomly with equal probabilities and used the population $N=10$. We run this genetic algorithm 300 times, and accumulated the database of lower-energy structures and their energies. Here, we assume that A atoms are larger and B atoms are smaller. The interactions between atoms $i$ and $j$ are given by the Lennard-Jones potentials:

$$v_{ij}(r) = 4\varepsilon_{ij}\left(\left(\frac{\sigma_{ij}}{r}\right)^{12} - \left(\frac{\sigma_{ij}}{r}\right)^{6}\right), \quad (1)$$

where the core radius $\sigma_{ij}$ represents the size of constituent atoms, and the potential well-depth $\varepsilon_{ij}$ represents the chemical bonding.

First, we only pay attention to the effect of the size difference on the lowest-energy structure of the 13-atom clusters. By changing the size ratio $\lambda=\sigma_{BB}/\sigma_{AA}$ (<1) and fixing the well-depth $\varepsilon_{AA}=\varepsilon_{BB}$, we can study the effect of size difference. The cross terms are determined from the Lorentz-Berthelot rule

$$\sigma_{AB} = \frac{\sigma_{AA} + \sigma_{BB}}{2}, \qquad \varepsilon_{AB} = \sqrt{\varepsilon_{AA}\varepsilon_{BB}} \quad . \quad (2)$$

A thorough study of the same system using molecular dynamics was reported by Cozzini and Ronchetti [16]. However, since they did not visualize the structures of the clusters in detail, we show the lowest- and the next-lowest-energy structures of $A_xB_{13-x}$ binary clusters when $\lambda=0.7$ in Figure 1. We see from Figure 1 that the lowest-energy structures are almost always icosahedral with one smaller atom (B) in the center of a distorted icosahedral cage (S-ICO) except for $A_{11}B_2$, which shows a structure with two smaller atoms in the center of a distorted non-icosahedral cage. We further notice that the tendency toward phase separation is clearly visible; the smaller atoms (B) tend to concentrate and squeeze into the center while the larger atoms (A) segregate on the surface of the cluster. This tendency is further exaggerated when the size ratio is smaller such as when $\lambda=0.6$. Then the lowest-energy structures become non-icosahedral.

Next we study the potential asymmetry in addition to the size difference using the Lennard-Jones potential designed by Kob and Andersen [20] which is artificially designed to form glass in the bulk. The potential parameters are:

$$\sigma_{AA}=1.0, \quad \sigma_{AB}=0.9, \quad \sigma_{BB}=0.88, \tag{3}$$

$$\varepsilon_{AA}=1.0, \quad \varepsilon_{AB}=1.5, \quad \varepsilon_{BB}=0.5,$$

which apparently violate the Lorentz-Berthelot rule (2).

Figure 2 shows the lowest- and the next-lowest-energy structures of the $A_xB_{13-x}$ Lennard-Jones binary clusters. The distorted icosahedral cluster is observed only for the $A_1B_{12}$ and $A_2B_{11}$ clusters and the next-lowest-energy structure of the $A_6B_7$ cluster. Furthermore, the center of icosahedra is a *larger* atom (A), which differs markedly from the structures shown in Figure 1. Several authors have noticed, however, that the lowest-energy structures of isolated binary clusters [16] as well as those in model binary glasses [15] are almost always distorted icosahedra with a smaller atom in the center of cage (S-ICO) rather than that with a larger atom in the center (L-ICO). Our results in Figure 2 seem to indicate that it is not always the general rule.

Finally we study the energy spectra of clusters in Figure 3. Several authors [16, 21] have noted that the lowest energy of a 13-atom cluster with icosahedral structure is isolated from other higher energies with non-icosahedral structures by a large energy gap. Thus the icosahedral cluster not only has the lowest energy but is thermally very stable [22] which can be seen in the energy spectra of $A_0B_{13}$, $A_{13}B_0$ and $A_1B_{12}$ clusters in Figure 3. Our results, however, show that the other lowest-energy non-icosahedral structures are not separated energetically from higher-energy metastable structures and are not thermally stable.

## 4. Discussion

When we pay attention only to the effect of size difference $\lambda$, our results together with those of Cozzini and Ronchetti [16] seem to indicate that the formation of an icosahedral cluster is favored only when the size ratio $\lambda$ is not too small (roughly $\lambda>0.6$-$0.7$). This size ratio $\lambda$ should be larger when the concentration of smaller atoms (B) is smaller. If the icosahedral cluster is related in some way or another to the formation of glass, the glass formation is also limited by these two conditions. Actually, similar criteria of the size ratio and the concentration for the glass formation were deduced empirically [23] and predicted from computer simulation [15, 24]. The tendency to the phase separation when $\lambda$ is small seems also in accord with the result of the computer simulation in [25].

When we look at the structures of the Lennard-Jones clusters with Kob and Andersen [20] potential parameters given by (3), the icosahedral structure is expected only when the smaller atoms are dominant in the cluster. So far as the authors know there seems to be no detailed study of the cluster structures in the Kob-Andersen glass in the bulk. Our total potential energy calculation for the Kob-Anderson clusters does not seem to support the statement that the icosahedral cluster is related to the glass formation. We know that the icosahedral clusters are observed in glass in many cases, which means that the formation of icosahedral clusters is a *sufficient condition* for glass formation. However, our calculation for the Kob-Andersen clusters seems to indicate that the formation of icosahedral clusters is *not* a *necessary condition* for glass formation.

## 5. Conclusion

In this paper, we have studied the lower-energy structures of the 13-atom binary Lennard-Jones clusters using the modified space-fixed genetic algorithm. When only the size difference exists, our result shows that the lowest-energy structures of 13-atom clusters are icosahedral when the size difference is not too large. This fact qualitatively explains the empirical rule that the moderate atomic size difference favors the glass transition [23] and also provides an evidence of the correlation between the existence of icosahedral clusters and the glass formation.

We have also studied the potential asymmetry in addition to the size difference using the Lennard-Jones potential used by Kob and Andersen. Curiously, the lowest-energy structures of 13-atom clusters are mostly non-icosahedral. This fact seems to contradict the statement that the formation and the stability of icosahedral clusters in close-packed liquids are related to the glass formation. There seem still much to be clarified in order to understand the microscopic mechanism of glass transition in binary alloys.

**Acknowledgement**

MI is partially supported from the grant for the promotion of advanced research in private institute by the ministry of education, sports, culture, science and technology of Japan.

**Figure Captions**

**Fig 1**. The lowest-energy (lower) and the next lowest-energy (upper) structures of Lennard-Jones $A_xB_{13-x}$ binary clusters with only a size asymmetry of $\lambda=0.7$. Except for $A_{11}B_2$, the lowest-energy structures are icosahedral. The phase separation within a cluster is visible. The figures show that the smaller atoms tend to concentrate and squeeze into the center of the cluster.

**Fig 2**. The lowest-energy (lower figure) and the next-lowest-energy (upper figure) structures of Lennard-Jones $A_xB_{13-x}$ binary clusters with Kob-Andersen parameters. Except for $A_1B_{12}$ and $A_2B_{11}$ the lowest-energy structures are non-icosahedral. In contrast to the case when only the size difference exists in Fig 1, the central atom of the cage is sometimes a larger atom rather than a smaller one. The phase separation is not visible within a cluster.

**Fig 3**. The energy spectra of clusters for all concentrations from $A_0B_{13}$ to $A_{13}B_0$ of the Lennard-Jones $A_xB_{13-x}$ binary clusters with Kob-Andersen parameters. The vertical scale represents the total potential energy of the cluster in the same unit as in eq. (3). The origin of the energy is shifted to the lowest-energy for each composition (Energy 0 is the lowest-energy). The horizontal scale represents the number of structures found within the energy window of the width 0.1. Therefore, there is always at least one energy level at energy 0. The spectra are not exhaustive since we run our genetic algorithm only 300 times with population $N$=10, which means we have used only 3000 samples. The lowest-energy structures of $B_{13}$ ($A_0B_{13}$) and $A_{13}$ ($A_{13}B_0$) are regular icosahedron, whose energy is separated from other higher energy structures by a large energy gap, while the other lowest-energy structures except for $A_1B_{12}$ are non-icosahedral whose energies are not separated from others energetically and will not be thermally stable.

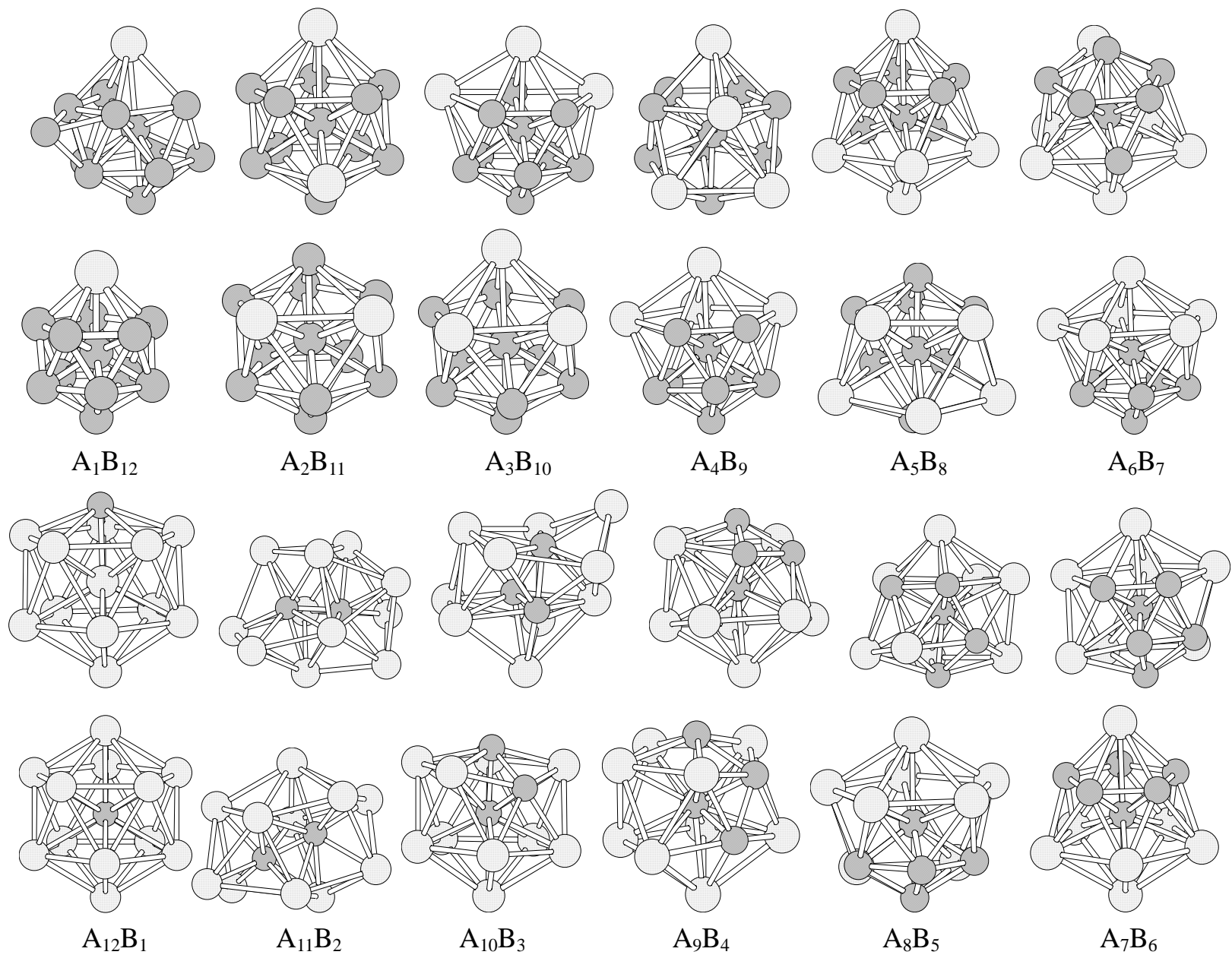

Fig. 1

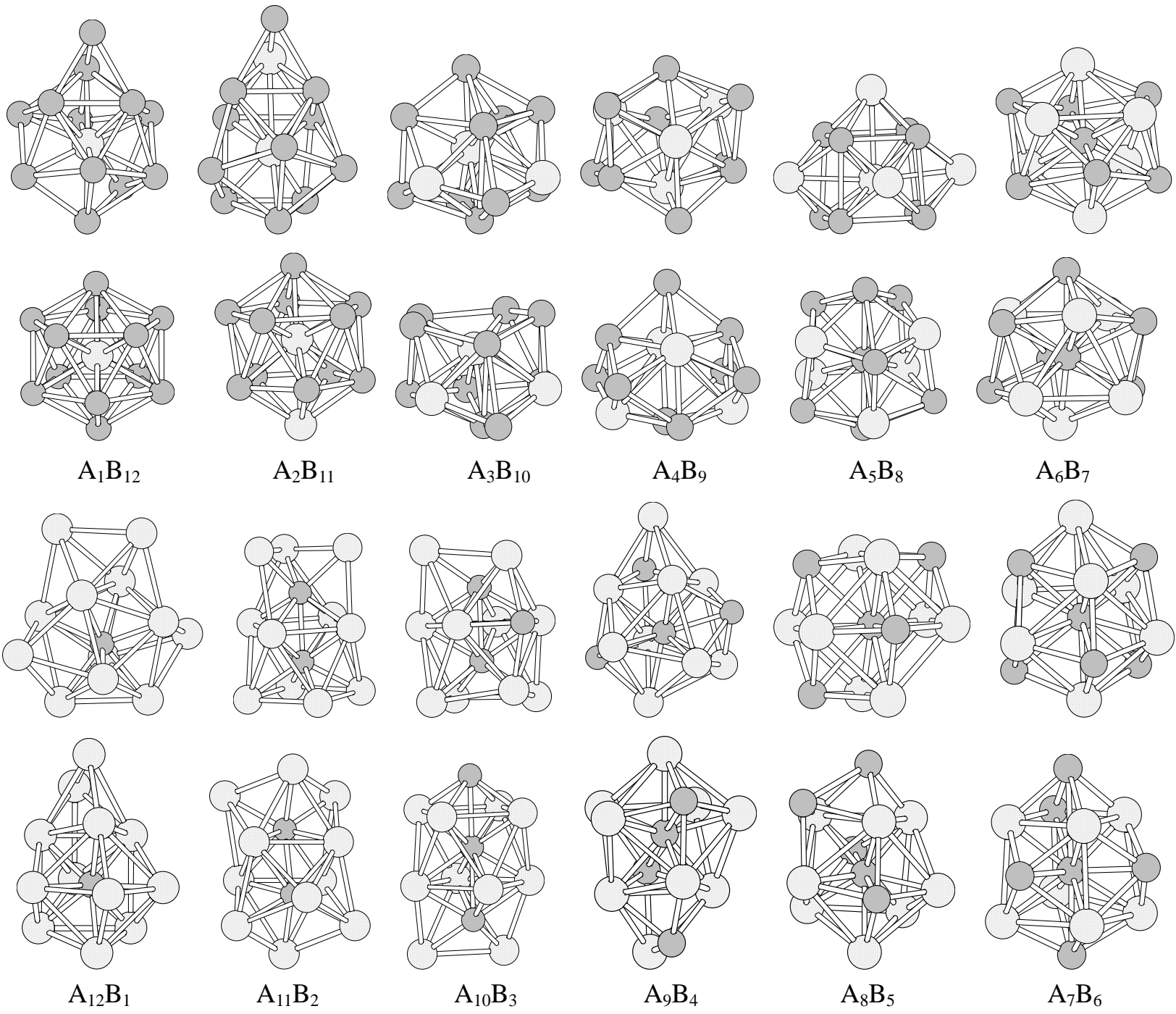

Fig.2

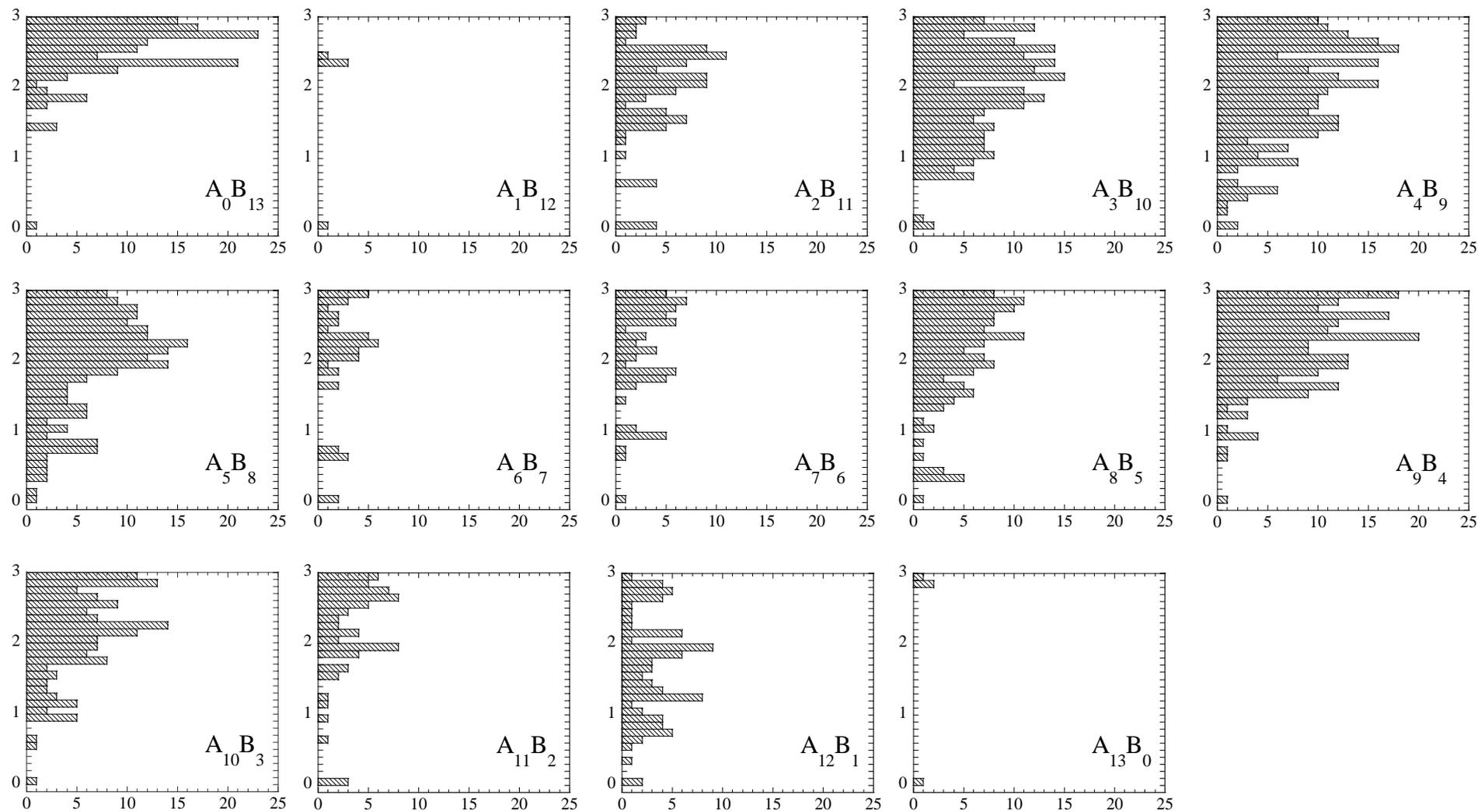

Fig.3